\documentclass[epsfig,12pt]{iopart}
\usepackage{epsfig}
\usepackage{iopams}

\begin{document}

\title[Polarizations for a gravitational wave in quadratic gravity]{Wave polarizations for a beam-like gravitational wave in quadratic
curvature gravity}

\author[de Rey Neto]{E. C. de Rey Neto, J.C.N. de
Araujo, O. D. Aguiar}

\address{Instituto Nacional de Pesquisas Espaciais - Divis\~ao de
Astrof\'\i sica \\ Av. dos Astronautas 1758, S\~ao Jos\'e dos
Campos, 12227-010 SP, Brazil}

\ead{\mailto{edgard@das.inpe.br},
     \mailto{jcarlos@das.inpe.br}, and
     \mailto{odylio@das.inpe.br}}

\begin{abstract}
We compute analytically the tidal field and polarizations of an exact
gravitational wave generated by a cylindrical beam of null matter of finite width and length in
quadratic curvature gravity. We propose that this wave can represent the
gravitational wave that keep up with the high energy photons produced in a gamma ray burst source. 
\end{abstract}

\submitto{\CQG}



\section{Tidal field in a spacetime of a $pp$-wave}
\label{sec2}

The relative accelerations between particles
located at nearby geodesics are determined by the geodesic
deviation equation:
\begin{equation}
\frac{D^2X^\mu }{d\tau^2}=-{R^\mu }_{\nu\gamma\delta}u^\nu  X^\gamma  u^\delta
,
\label{geod1}
\end{equation}
where $\tau$ is the proper time of one of the particles and the right hand side represents the tidal force.
By choosing a orthonormal tetrad~$\{\mbox{\boldmath${e}$}_{\hat{a}}\}$ such that
$\mbox{\boldmath${e}$}_{\hat{0}}=\mbox{\boldmath${u}$}$ is the four-velocity of one of the test particles and
$\{\mbox{\boldmath${e}$}_{\hat{i}}\}$, $i=1,2,3$ are orthogonal space-like unit four-vectors,
such that $\mbox{\boldmath${e}$}_{\hat{a}}\cdot\mbox{\boldmath${e}$}_{\hat{b}}
\equiv
g_{\mu\nu}e^\mu_{\hat{a}}e^\nu_{\hat{b}}=\eta_{\hat{a}\hat{b}}={\rm
diag}(-1,1,1,1)$, we obtain that $\ddot{X}^{\hat{0}}=R_{\mu\nu\gamma\delta}u^\mu  u^\nu  X^\gamma
u^\delta=0$ and
\begin{equation}
\ddot{X}^{\hat{i}}=-{R^{\hat{i}}}_{\hat{0}\hat{j}\hat{0}}X^{\hat{j
}},
\label{accspace}
\end{equation}
where the overdots means derivatives with respect to $\tau$ and ${R^{\hat{i}}}_{\hat{0}\hat{j}\hat{0}}$ are the projections of the Riemann
tensor components on the tetrad frame~$\{\mbox{\boldmath${e}$}_{\hat{a}}\}$.

Consider a gravitational  $pp$-wave given by~\cite{KSMH}:
\begin{equation}
ds^2=-dudv+H(u,r,\phi)du^2+dr^2+r^2d\phi^2,
\label{metric}
\end{equation}
where ($\hbar=c=1$),  $u=t-z$, $v=t+z$, $r$ and $\phi$ polar coordinates in the plane
perpendicular to the wave propagation direction. The non vanishing accelerations are given by
\numparts
\begin{eqnarray}
\ddot{X}^{\hat{1}}=-(A_++A_\circ)X^{\hat{1}}+A_\times X^{\hat{2}},\\
\ddot{X}^{\hat{2}}=+A_\times X^{\hat{1}}-(-A_++A_\circ)X^{\hat{2}},
\label{eqgeodev}
\end{eqnarray}
\endnumparts
where
\begin{equation}
\eqalign
{
A_\circ=\frac{1}{8}\nabla^2_\perp H
\label{phi22},\\
A_+=\!\frac{1}{8}\!\left(\!\frac{H_{,\phi\phi}}{r^2}\!+\!\frac{H_{,r
}}{r}-H_{,rr}\!\right)\qquad,\qquad
\label{A+1}
A_\times=\!\frac{1}{4}\!\left(\frac{H_{,r\phi}}{r}-\frac{H_{,\phi}}{r
^2}\right)\!.
}
\label{Ax1}
\end{equation}
The comma stands for partial derivatives and $\nabla_\perp^2$ is the Laplacian in the transverse plane~\cite{Rsub}. All patterns are transverse.
The quantities $A_+$ and $A_\times$ generate helicity-2 polarization patterns shifted by $\pi/4$ while $A_\circ$ produces an helicity-0 pattern. 
We apply the equations~(4) only at large distances from the massive radiating bodies.

\section{Gravitational wave generated by a  cylindrical beam of photons}

\label{sec1}
\setcounter{figure}{0}
\begin{figure}
\begin{center}
\epsfig{figure=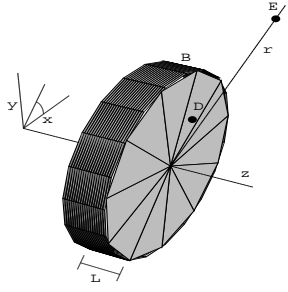,angle=360,height=5cm,width=6cm}
\caption{\label{fig1}A cylindrical beam of radius $B$ and length $L$ of high energy photons propagating with light
velocity in the $z$ direction. An observer {D} is crossed by
the photons beam and an observer {E} lies outside  the beam.}
\end{center} 
\end{figure}

The quadratic gravity equations for the spacetime~(\ref{metric}) becomes:
\begin{equation}
-\frac{1}{2}[\beta\nabla_\perp^4+\nabla_\perp^2]H(u,r,\phi)=8\pi G T_{uu},
\label{qeq}
\end{equation}
where $G$ is the Newton's gravitational constant and $\beta$ the coupling parameter of the Ricci squared term in the gravitational action~\cite{buchdahl}.
Consider a cylindrical beam of photons travelling in the $z$ direction with constant energy density $\varrho_0$, radius $B$ and length ${\rm L}=c(t_{F}-t_{I})$, where $t_{F}-t_{I}\equiv\delta t>0$ is burst duration~(FIG.~\ref{fig1}), we have
\begin{equation}
T_{uu}=\varrho_0\Theta(u-u_{I})\Theta(u_{F}-u)\Theta(B-r),
\label{Tuu}
\end{equation}
where $\Theta(x)$ is the Heaveside step function, $u_{I}=t_{I}-z$, $u_{F}=t_{F}-z$ and $\varrho_0$ is the energy density of the beam.

The solution of~(\ref{qeq}) with the source~(\ref{Tuu}) is given by:
\begin{equation}
H(x,u)=h(x)f(u),
\label{H(rho,u)}
\end{equation}
where $f(u)$ is $1$ if $u_{I}<u<u_{F}$ and 0 otherwise,
\begin{eqnarray}
h(x)&=&\kappa\{[4BbK_1(b_0)I_0(x)-r^2-4b^2]\Theta[b(b_0-x)]\nonumber \\
&-&[2B^2\ln(x)+B^2+4BbI_1(b_0)K_0(x)]\Theta[b(x-B_0)]\},
\end{eqnarray}
$\kappa=4\pi G\varrho_0$, $x\equiv r/b$, $b_0=B/b$, $K_\nu$ and $I_\nu$ are modified Bessel functions and $b\equiv(-\beta)^{(1/2)}$~\cite{RAA1}. We assume that $\beta<0$ since for $\beta>0$ it is known that there is no acceptable Newtonian limit for the nonrelativistic gravitational potential between point masses. The solution~(\ref{H(rho,u)}) implies that spacetime is flat for $u<u_{I}$ and $u>u_{F}$ and curved for $u_{I}<u<u_{F}$.
The quantities~(\ref{phi22}) becomes:
\begin{eqnarray}
A_\circ=-&\frac{\kappa}{2}&\{[b_0K_1(b_0)I_0(x)-1]\Theta[b(b_0\!-\!x)]\nonumber \\
&-&b_0I_1(b_0)K_0(x)\Theta[b(x-b_0)]\}f(u),
\label{A_0}
\end{eqnarray}
\begin{eqnarray}
A_{+}=&-&\frac{\kappa}{2}\{b_0K_1(b_0)I_2(x)\Theta[b(b_0\!-\!x)]+[b_0^2/x^2\nonumber \\
&-&b_0I_1(b_0)K_2(x)]\Theta[b(x-b_0)]\}f(u)
\label{A+}
\end{eqnarray}
and $A_\times=0$, since $H_{,\phi}=0$.
\begin{figure}
\begin{center}
\begin{tabular}{c@{\qquad}c}
\mbox{\epsfig{figure={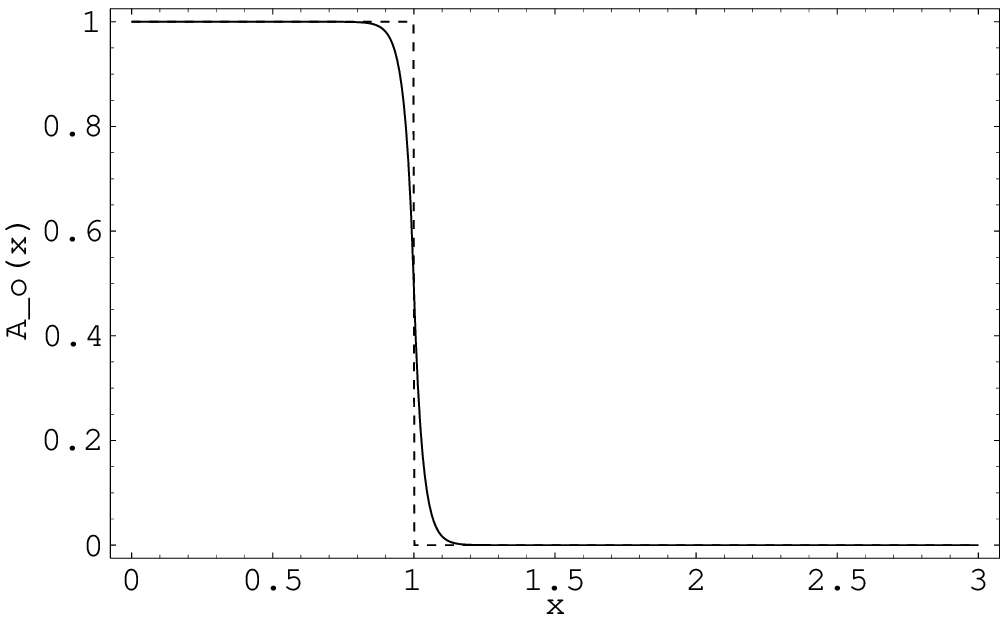},angle=360,height=3cm,width=5cm}} &
\mbox{\epsfig{figure=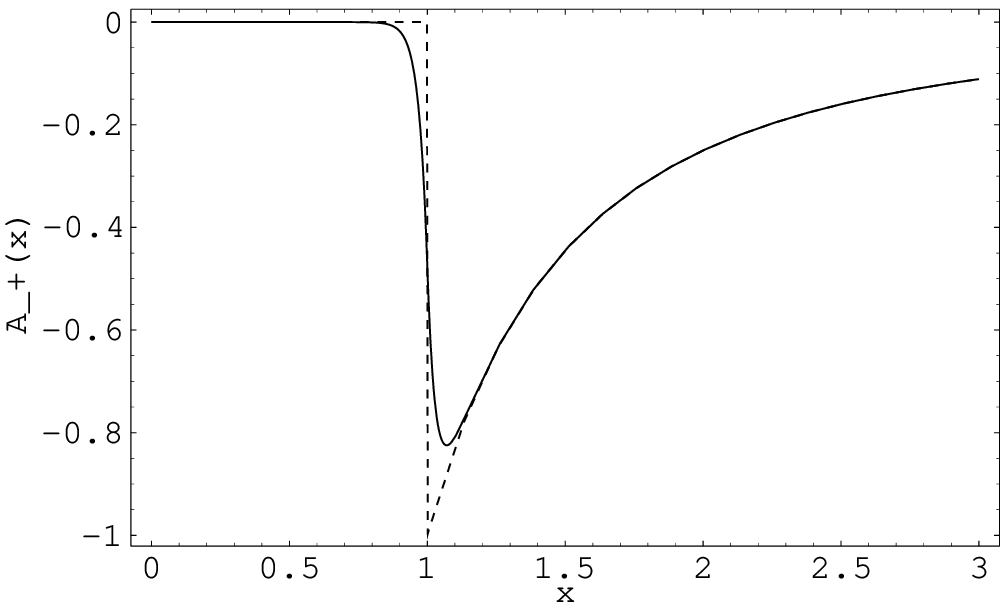,angle=360,height=3cm,width=5cm}}\\
\hbox{(a)}    &  \hbox{(b)}                  \\
\end{tabular}
\caption{\label{fig2} The profiles of $A_\circ$ and $A_+$. The solid curves represent the solutions to quadratic gravity and the dashed one represents the solutions to Einstein's gravity. The units are such that $\kappa/2=1$ and $x\rightarrow x/b_0$. We set a large value of $b$ to obtain a better visualization of the quadratic curvature effect.}
\end{center}
\end{figure}
In~(FIG.~\ref{fig2}) we compare the profiles of $A_0$ and $A_+$ as functions of $x$ in the Einstein's and quadratic gravity. 

For a distant GRB progenitor, we can roughly approximate the energy density by
\begin{equation}
 \varrho_0\simeq \frac{E}{4\pi z^2c\delta t},
 \label{Eeq}
 \end{equation}
where $z$ is the distance to the (GRB) source and $E$ is the burst energy. 
If $r<B$, $T_{uu}=\varrho_0 f(u)$; there is a radiating field of non gravitational energy.  If $r>B$, there are no radiating fields and $T_{uu}=0$. 

\section{Effect on geodetic test particles}
Consider observers D and E which measures the relative accelerations between test particles at a great distances form the GRB progenitor, such that $\varrho_0$ is given by~(\ref{Eeq}). The region ($r<B$; $x<b_0$) is not a pure vacuum since, $T_{uu}=\varrho_0f(u)$. Therefore, there is a helicity 0 polarization pattern in addition to the helicity 2~\cite{K,PO} :
\begin{equation}
A_\circ\!=\!\frac{\kappa}{2}[1-b_0K_1(b_0)I_0(x)]f(u)\;\; {\rm and}\;\; A_+\!=\!-\frac{\kappa}{2}b_0K_1(b_0)I_2(x)f(u).
\label{A0A+D}
\end{equation}
The quadratic gravity contributions are negligible with respect to the Einstein's gravity ones unless $r\simeq B$.

For the observers {E}, ($r>B$; $x<b_0$), 
\begin{equation}
A_\circ=\frac{\kappa}{2}b_0 I_1(b_0)K_0(x)f(u)\;\; {\rm and}\;\; A_+=\!-\frac{\kappa}{2}\left[\frac{b_0^2}{x^2}+b_0I_1(b_0)K_2(x)\right]\!f(u).
\label{Eobs}
\end{equation}
Here, the $A_\circ$ comes only from quadratic curvature interactions because there are no radiating fields in this region. 
We must not worry about the appearance of the helicity 0 component in~(\ref{A0A+D}), since this is the expected result when radiating fields are present~\cite{K,PO}.

The greatest amplitude of the effect of the gravitational wave on geodetic test particles near the Earth is proportional to $2\pi G\varrho_0\delta t^2\sim 10^{-38}$ (in dimensionless units) for a flux of $\sim 10^{-2}$ erg ${\rm cm}^{-2}{\rm s}^{-1}$ and $\delta t~\sim 10$s typical values for a GRB.

\section{Summary and Conclusions}

We compute an exact $pp$-wave solution to quadratic gravity equations with a cylindrical beam of photons as a source. We propose that this model can represent approximately a gravitational wave accompaning a beam of high energy photons emitted in a GRB. By considering the geodesic deviations far from the GRB source we show that, for an observer that is crossed by the beam ($r<R$) during the interval $\delta t$, the helicity-2 polarization pattern is given only by the quadratic curvature effects and is negligible unless the observer is located at $r\simeq B$. This observer must see an helicity-0 pattern in the relative accelerations of test particles even in GR gravity. This result do not conflict with the GR expectations, since this observer is crossed by the photons beam and GW at the same time.  An observer that is not crossed by the beam sees only a helicity-2 pattern which decreases with the square of the distance form the beam axis.
The magnitude of the effect of the (GW) produced by the photons beam at the Earth on geodetic test particles is obviously very small.

\ack{We want to thank the Brazilian agency {FAPESP} for the financial
support (grants: 00/10374-5, 98/13468-9, 03/04342-1) and CNPq (grants:
300619/92-8, 304.666/02-5).}

\end{document}